\documentclass[a4paper,11pt]{article}
\pdfoutput=1

\usepackage[utf8]{inputenc}
\usepackage[T1]{fontenc}
\usepackage[english]{babel}
\usepackage{layout}
\usepackage{setspace}
\usepackage{lmodern}
\usepackage{enumitem}
\usepackage{soul}
\usepackage{eurosym}

\usepackage{verbatim}
\usepackage{moreverb}

\usepackage{sectsty}

\usepackage{mathrsfs}
\usepackage{cancel}

\usepackage{makeidx}
\usepackage{multirow}
\usepackage{stmaryrd}

\usepackage{docstyle}

\title{\textcolor{mycolor}{Exorcizing Ghosts from the Vacuum Spectra\\in String-Inspired Nonlocal Tachyon Condensation}}

\author[a,b]{Florian Nortier}

\affiliation[a]{Université Paris-Saclay, CNRS, CEA, Institut de Physique Théorique,\\ 91191, Gif-sur-Yvette, France}

\affiliation[b]{Université Claude Bernard Lyon 1, CNRS/IN2P3, IP2I Lyon, UMR 5822,\\ Villeurbanne, F-69100, France}

\emailAdd{f.nortier@ip2i.in2p3.fr}

\abstract{Tachyon condensation in quantum field theory (QFT) plays a central role in models of fundamental interactions and cosmology. Inspired by tower truncation in string field theory, ultraviolet completions were proposed with infinite-derivative form factors that preclude the appearance of pathological ghosts in the particle spectrum, contrary to other local higher-derivative QFT's. However, if the infinite-derivative QFT exhibits other vacua, each of them has its own spectrum, which is generally not ghost-free: an infinite tower of ghost-like resonances pops up above the nonlocal scale at tree-level, whose consistency is unclear. In this article, a new weakly nonlocal deformation of a generic local QFT is introduced via a Lorentz and gauge covariant star-product of fields, which is commutative but nonassociative in general. This framework realizes tachyon condensation without ghosts at the perturbative level, with applications for spontaneous symmetry breaking.}


\arxivnumber{2307.11741}

\begin{document}

\maketitle
\flushbottom

\section{Introduction}
\label{introduction}
Despite the incredible success of effective field theories (EFT's) \cite{Burgess:2020tbq} as local quantum field theories (QFT's) \cite{Peskin:1995ev} to describe the phenomenology of elementary particles \cite{Workman:2022ynf} and primordial cosmology \cite{Baumann:2018muz}, it is usually expected that the fundamental quantum theory, which embeds gravity with the subatomic interactions, should be nonlocal \cite{Marshakov:2002ff, Giddings:2011xs}. The most popular example is string theory \cite{Becker:2007zj}, where the perturbative degrees of freedom are not pointlike particles anymore but fundamental extended objects: strings.

String theory is traditionally defined perturbatively in the worldsheet formulation \cite{Becker:2007zj}. Among the nonperturbative approaches, string field theory (SFT) \cite{Erbin:2021smf} is an off-shell QFT formulation: a string field creates and annihilates strings, whose interaction vertices are manifestly nonlocal. Such a string field can be expanded into an infinite tower of particle fields with increasing spins, which are interpreted as string excitations. Cubic SFT has been used to study D-brane decay by tachyon condensation: the tachyon of open bosonic strings attached to a D-brane rolls down towards the asymptotic minimum of the potential, and the brane decays into a lower-dimensional brane or the closed string vacuum \cite{Ohmori:2001am}. This phenomenon can be studied in a simplified framework, where the string tower is cut at 0-level: one gets a tachyon field with a weakly nonlocal (WNL) cubic potential, exhibiting an infinite-derivative form factor \cite{Efimov:1968flw, Namsrai:1986md, Moeller:2002vx, Vladimirov:2003kg, Barnaby:2007ve, Calcagni:2007ef, Barnaby:2008tc, Calcagni:2009jb, Barnaby:2010kx, Gorka:2012hs, Tomboulis:2015gfa, Cembranos:2016dll, Dragovich:2017kge, Calcagni:2018lyd, Boos:2020qgg, Erbin:2021hkf, BasiBeneito:2022wux, Heredia:2023cgs}, i.e. a pseudodifferential operator.

\begin{figure}[h!]
\begin{center}
\includegraphics[height=10cm]{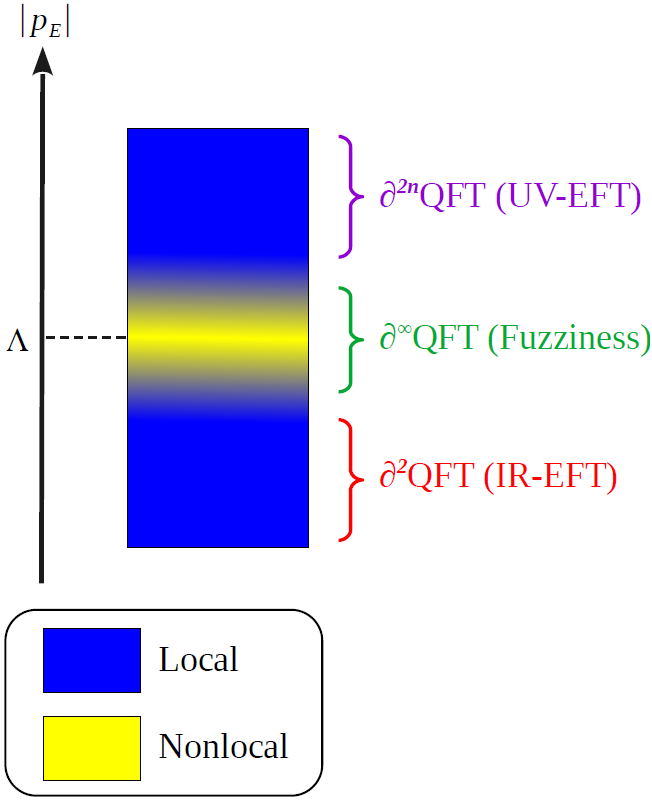}
\end{center}
\caption{\textit{Sketch of UV-completing an EFT through an $\partial^\infty$QFT with a Kuz'min-Tomboulis form factor. The theory interpolates between a local 2-derivative EFT at low-energy (IR) and a local higher-derivative EFT at high-energy (UV) via an infinite-derivative form factor. A nonlocal window around the scale $\Lambda$ replaces the pathological ghosts of local higher-derivative QFT's. The proof of locality of counter-terms in the WNL window has not been addressed in a complete model \cite{Tomboulis:1997gg}.}}
\label{energies-fuzzy}
\end{figure}

The study of string-inspired nonlocal field theories has triggered a revival of the old program\footnote{For a review of the pre-80s literature on infinite-derivative QFT, cf. Ref.~\cite{Namsrai:1986md}.} to improve the ultraviolet (UV) behavior in QFT by some WNL deformation. Leaving aside SFT, if one considers an infinite-derivative quantum field theory ($\partial^\infty$QFT) with only interacting Klein-Gordon scalars and Dirac fermions, one can delocalize the interaction terms, and the theory becomes UV-finite \cite{Namsrai:1986md, Buoninfante:2018mre}. Perturbative analyticity and unitarity\footnote{Technically, these properties hold if the $\partial^\infty$QFT is defined by Efimov analytic continuation \cite{Efimov:1966ylf}, sometimes (badly) called generalized Wick rotation, even if there is no rotation of the integration contour \cite{Erbin:2021smf}: the path integral is defined in Euclidean signature, and the Minkowskian $\partial^\infty$QFT is obtained by analytic continuation of the external momenta. The reason behind this subtlety is an essential singularity at complex infinity $\widetilde{\infty}$ that forbids the (true) Wick rotation.} have been shown for both SFT \cite{Pius:2016jsl, Sen:2016bwe, Sen:2016gqt, Sen:2016uzq, Sen:2016ubf, Pius:2018crk, DeLacroix:2018arq, Erbin:2021smf} and these simpler $\partial^\infty$QFT's \cite{Efimov:1966ylf, Carone:2016eyp, Briscese:2018oyx, Christodoulou:2018jbn, Chin:2018puw, Briscese:2021mob, Koshelev:2021orf, Buoninfante:2022krn}. The case of (non-SFT) $\partial^\infty$QFT's with gauge symmetries is more delicate since (local) gauge transformations impose a competition between WNL form factors in the kinetic and interaction terms \cite{Chretien:1954we, Evens:1990wf, Kleppe:1991rv}. If one keeps locality in gauge transformations, the Krasnikov-Terning scheme \cite{Krasnikov:1987yj, Terning:1991yt, Biswas:2011ar} is manifestly gauge invariant: several authors argued that one can build WNL extensions of Yang-Mills theories or Einstein gravity that are superrenormalizable or UV-finite by power counting thanks to the special class of Kuz'min-Tomboulis (KT) form factors \cite{Kuzmin:1989sp, Tomboulis:1997gg, Modesto:2011kw, Modesto:2014lga, Modesto:2015lna, Modesto:2015foa, Tomboulis:2015esa, Giaccari:2016kzy, Modesto:2016max, Modesto:2017hzl, Modesto:2017sdr, Koshelev:2017ebj, BasiBeneito:2022wux}, i.e. WNL form factors that are asymptotically local in the deep-UV (cf. Fig.~\ref{energies-fuzzy}). This framework may offer a potential alternative path to string theory towards a fundamental theory of Nature that is entirely based on QFT.

One may wonder if an $\partial^\infty$QFT is not plagued by the infamous ghost problem of usual higher-derivative QFT's (cf. Refs.~\cite{Woodard:2015zca, Platania:2022gtt} for reviews). However, the infinite-derivative form factors are carefully chosen to not introduce new singularities in the complex plane beyond what one already has in the healthy local limit of the theory; this has been checked in perturbation theory \cite{Pius:2016jsl, Sen:2016bwe, Sen:2016gqt, Sen:2016uzq, Sen:2016ubf, Pius:2018crk, DeLacroix:2018arq, Erbin:2021smf, Efimov:1966ylf, Carone:2016eyp, Briscese:2018oyx, Chin:2018puw, Briscese:2021mob, Koshelev:2021orf, Buoninfante:2022krn}. Indeed, contrary to early expectations \cite{Eliezer:1989cr}, the ghosts (that one naively guesses from truncating the infinite-derivative form factor at some finite derivative degree) decouple once the whole form factor is resumed \cite{Namsrai:1986md, Boos:2021chb, Boos:2021jih, Boos:2021lsj, Boos:2022biz}, as it should when one wants to probe the WNL scale. Instead of ghosts that propagate backward in time and interact with acausal local vertices \cite{Coleman:1969xz, Grinstein:2008bg, Alvarez:2009af, Donoghue:2019ecz}, weak nonlocality (WNL) introduces some fuzziness in spacetime resolution by smearing the vertices \cite{Efimov:1968flw, Namsrai:1986md, Barnaby:2010kx, Tomboulis:2015gfa, Keltner:2015xda, Carone:2016eyp, Buoninfante:2018mre, Giaccari:2018nzr, Boos:2020qgg}.

In particle physics, tachyon condensation is used for spontaneous symmetry breaking (SSB). In the standard model (SM), it is at the heart of the so-called Higgs mechanism \cite{Nambu:1960tm, Anderson:1963pc, Higgs:1964ia, Englert:1964et, Higgs:1964pj, Guralnik:1964eu, Higgs:1966ev, Migdal:1966tq, Kibble:1967sv, tHooft:1971qjg, Guralnik:2011zz}: the cornerstone of the electroweak theory \cite{Glashow:1961tr, Salam:1968rm, Weinberg:1967tq, Paschos:2007pi}. Naturally, WNL is expected to provide a way to stabilize the scalar potential under radiative corrections \cite{Krasnikov:1987yj, Moffat:1990jj, Biswas:2014yia, Ghoshal:2017egr, Frasca:2020jbe, Nortier:2021six}. String-inspired scalar fields with nontrivial potentials have also been studied in cosmology\footnote{WNL gravity has applications in black hole physics and cosmology, cf. Refs.~\cite{Buoninfante:2022ild, Koshelev:2023elc} for reviews.}, with applications for inflation, dark energy, phase transition/vacuum decay, etc \cite{Cline:2002it, Arefeva:2004odl, Arefeva:2005mom, Arefeva:2006rnj, Barnaby:2006hi, Cai:2007gs, Koshelev:2007fi, Arefeva:2007wvo, Lidsey:2007wa, Barnaby:2007yb, Calcagni:2007ru, Joukovskaya:2007nq, Lidsey:2008zz, Dimitrijevic:2008ni, Arefeva:2008zru, Mulryne:2008iq, Joukovskaya:2008cr, Barnaby:2008pt, Koshelev:2009ty, Calcagni:2009dg, Vernov:2009vf, Biswas:2009nx, Calcagni:2010ab, Vernov:2010ui, Biswas:2010xq, Vernov:2010ft, Koshelev:2010bf, Galli:2010ti, Galli:2010qx, Arefeva:2011qvf, Arefeva:2011yua, Biswas:2012ka, Chialva:2014rla, Koshelev:2016vhi, Milosevic:2016utg, Sheikhahmadi:2020prj, Koshelev:2020fok, Dragovich:2022vvh, Frasca:2022vvp, Ghoshal:2022mnj, Sadeghi:2023cxh}. Nevertheless, such UV-completion is nontrivial because a string-inspired tachyonic potential is usually ghost-free in only 1 vacuum of the potential while, in the other vacua, the spectrum develops an infinite tower of ghosts whose consistency is controversial \cite{Barnaby:2007ve, Barnaby:2008pt, Galli:2010qx, Barnaby:2010kx, Gama:2018cda, Hashi:2018kag, Koshelev:2020fok}. Any attempt to embed local models with tachyon condensation in an $\partial^\infty$QFT that is UV-complete should address this ghost issue.

In the literature, only 2 solutions in a non-SFT framework have been proposed to the ghost problem in the infinite-derivative Higgs mechanism:
\begin{itemize}[label=$\spadesuit$]
\item \textit{Gauge-Higgs Unification:} In Ref.~\cite{Hashi:2018kag}, the authors added 2 extra dimensions of space to identify the Higgs field with the extra components of 6D gauge fields, such that they avoid the ghost-like tower at tree-level.
\item \textit{Tree-Duality:} In Ref.~\cite{Modesto:2021ief}, Modesto proposed a recipe to generate an $\partial^\infty$QFT from a local QFT by imposing that the perturbative spectra and scattering amplitudes match at tree-level \cite{Modesto:2021soh}, while keeping superrenormalizability and UV-finiteness by power counting at loop-level\footnote{The proof of locality of counter-terms has only been addressed in the asymptotic UV-regime.} \cite{Modesto:2022asj, Calcagni:2023goc}. As a byproduct, this ``tree-duality'' recipe allows building a ghost-free Higgs mechanism at tree-level \cite{Modesto:2021okr}.
\end{itemize}
The aim of the present study is to propose a new WNL deformation that realizes tachyon condensation without vacuum instabilities at the tree-level.

This article is organized as follows. In Section~\ref{section_problem}, the problem of vacuum stability induced by ghosts in string-inspired tachyon condensation is presented in detail. In Section~\ref{WNL_deformation}, a new solution to this issue is proposed and applied for SSB of discrete and gauge symmetries (Higgs mechanism). Appendix~\ref{conventions} gives a list of conventions and notations.

\section{Infinite-Derivatives vs. Vacuum Stability}
\label{section_problem}

\subsection{Monster Zoology: Tachyons, Ghosts \& Co.}
Consider the Minkowskian tree-level propagator for a particle of (complex) mass $m$ in QFT, with the $i \epsilon$ prescription that preserves unitarity:
\begin{equation}
\dfrac{i c_{gh} \, R}{p^2 - c_{th} \, m^2 +  i \epsilon} \, ,
\label{propa_0}
\end{equation}
where: $c_{gh} \, R$ is the residue of the complete propagator at the corresponding pole that is real positive for real poles and complex for complex conjugate poles; $c_{th}$ is the sign of the mass term in the potential. In the following, one considers only particles defined by simple poles (i.e. of multiplicity 1). One can recast a propagator into a sum of such single particle propagators by performing a partial fraction decomposition:
\begin{equation}
\prod_{i=1}^N \dfrac{1}{p^2-m_i^2} = \sum_{i=1}^N \left( \sum_{j \neq i} \dfrac{1}{m_i^2-m_j^2} \right) \dfrac{1}{p^2-m_i^2} \, .
\end{equation}
In order to fix the vocabulary, one will follow Ref.~\cite{Platania:2022gtt}. One has: a \textit{canonical particle} for $c_{gh} = c_{th} = 1$; a \textit{ghost} for $c_{gh} = -c_{th} = -1$; a \textit{tachyon} for $c_{gh} = -c_{th} = 1$; a \textit{tachyon-ghost} for $c_{gh} = c_{th} = -1$. One can refer to Refs.~\cite{Cline:2003gs, Woodard:2015zca, Platania:2022gtt} for more details on the pathologies induced by ghosts or tachyons, i.e. vacuum instability and causality violation. Concerning analyticity in the $p_0^2$-complex plane, the pole of the propagator \eqref{propa_0} is located at
\begin{equation}
p_0^2 = E^2 - i \, \Gamma \, , \ \ \ 
E^2 = \vec{p}^{\ 2} + c_{th} \, \mathfrak{R}(m^2) \, , \ \ \ 
\Gamma = c_{th} \, \mathfrak{I}(m^2) + \epsilon \, .
\end{equation}
One can distinguish 3 kinds of poles:
\begin{itemize}[label=$\spadesuit$]
\item Canonical particles or tachyons have $\mathfrak{I}(m^2) = 0$ and $\epsilon > 0$. The poles lie in the $4^\text{th}$ quadrant of the $p_0^2$-complex plane.
\item Ghosts or tachyon-ghosts have $\mathfrak{I}(m^2) = 0$ and $\epsilon < 0$. The poles lie in the $1^\text{st}$ quadrant of the $p_0^2$-complex plane.
\item Complex conjugate poles, usually called Lee-Wick (LW) pairs \cite{Lee:1969fy, Cutkosky:1969fq, Lee:1970iw, Grinstein:2007mp, Wise:2009mi}, have $\mathfrak{I}(m^2) \neq 0$ and $\epsilon = 0$. The poles lie in the $1^\text{st}$ and $4^\text{th}$ quadrants of the $p_0^2$-complex plane.
\end{itemize}

These last years, the so-called fakeon (or Anselmi-Piva) prescription to treat ghosts and LW-pairs has been developed \cite{Anselmi:2017yux, Anselmi:2017lia, Anselmi:2017ygm, Anselmi:2018kgz, Anselmi:2021hab, Liu:2022gun, Anselmi:2022qor}, and recently generalized to continua of ghost quasiparticles associated to branch cuts \cite{Calcagni:2022shb}: it involves a nonanalytic Wick rotation from Euclidean to Minkowskian momenta. The spurious singularities turn to be purely virtual particles (fakeons) that cannot go on-shell. However, Kubo and Kugo recently claimed that the usual Feynman rules used in the literature (including the fakeon prescription) ignore some subtleties that arise with Dirac $\delta$-functions defined on $\mathbb{C}$, implying the existence of new contributions to the amplitudes that violate unitarity around the ghost threshold \cite{Kubo:2023lpz}. Therefore, there is no clear consensus in the literature on the issue of building a physically meaningful theory with ghosts. In this article, I will adopt the standard point of view in the QFT community: one has to impose analyticity on the physical sheet\footnote{Analyticity alone on the physical sheet allows essential singularities at complex infinity $\widetilde{\infty}$, e.g. the exponential function $e^z$ with $z \in \mathbb{C}$. Exploring this unusual class of theories is the raison d'être of the $\partial^\infty$QFT program.} \cite{Eden:1966dnq}. This excludes ghosts and LW-pairs that are considered to spoil\footnote{I apologize to the reader who disagrees with this statement.} either stability \cite{Cline:2003gs, Woodard:2015zca, Platania:2022gtt} or unitarity \cite{Kubo:2023lpz}.

\subsection{Anatomy of Vacua in Tachyon Condensation}
\label{vacua_tachyons}
Consider a local QFT defined on $\mathbb{R}^{1,3}$ with 1 real scalar field $\phi(x)$ and a $\mathbb{Z}_2$ symmetry, i.e. an invariance under $\phi(x) \mapsto - \phi(x)$. The Lagrangian is
\begin{equation}
\mathcal{L}_\phi = - \dfrac{1}{2} \,  \phi \, \square \phi - V (\phi) \, ,
\label{L_phi_1}
\end{equation}
where $V(\phi)$ is a potential with a quartic self-interaction\footnote{Motivated by the 0-level truncation of cubic SFT, the spectrum with a WNL cubic self-interaction was studied in Ref.~\cite{Hashi:2018kag}, with similar conclusions. In this article, the choice of a quartic self-interaction allows for highlighting the deep link with SSB.} $(M^2 \in \mathbb{R}, \ \lambda>0)$:
\begin{equation}
V(\phi) = \dfrac{M^2}{2} \, \phi^2 + \dfrac{\lambda}{4!} \, \phi^4 \, .
\label{V_phi_1}
\end{equation}
In the literature, WNL is usually introduced by SFT-like smeared interaction terms \cite{Buoninfante:2018mre}, obtained by performing the replacement $\phi^4(x) \mapsto \varphi^4(x)$ in $V(\phi)$. It involves the smeared field\footnote{This kind of smeared interaction $\varphi^4(x)$ is directly inspired by the cubic term in open bosonic SFT (e.g. Ref.~\cite{Calcagni:2009jb}). Note that the definition of $\varphi(x)$ on a Minkowski spacetime is ``symbolic''. A WNL field theory is defined via the path integral in Euclidean signature, and then the Green functions are analytically continued to a Lorentzian signature \cite{Efimov:1966ylf, Pius:2016jsl, Sen:2016bwe, Sen:2016gqt, Sen:2016uzq, Sen:2016ubf, Pius:2018crk, DeLacroix:2018arq, Erbin:2021smf, Carone:2016eyp, Briscese:2018oyx, Chin:2018puw, Briscese:2021mob, Koshelev:2021orf, Buoninfante:2022krn}.} $\varphi(x)$, with the WNL scale $\Lambda = \sqrt{1/\eta}$, given by
\begin{equation}
\varphi(x) = e^{\frac{1}{2} \, \vartheta \left( - \eta \, \square \right)} \, \phi(x) \, ,
\label{field_redef}
\end{equation}
that is defined by a pseudodifferential operator. $\vartheta(z)$ is an entire function on the complex plane, that is real on the real axis, and satisfying $\vartheta(0)=0$. The real parameter $\eta$ is a fundamental area that one will call ``fuzzy plaquette'', which characterizes the smearing of the WNL field interactions. Note that the resulting $\partial^\infty$QFT is still $\mathbb{Z}_2$-invariant, like the original local QFT.\\

Since the WNL form factor $e^{\vartheta(z)}$ has no zeroes on $\mathbb{C}$, one can perform a field redefinition $\phi(x) \mapsto \varphi(x)$, and the Lagrangian can be rewritten only in terms of the smeared field $\varphi(x)$ as
\begin{equation}
\mathcal{L}_\varphi = - \dfrac{1}{2} \, \varphi \, e^{-\vartheta \left( -\eta \, \square \right)} \left( \square + M^2 \right) \varphi - \dfrac{\lambda}{4!} \, \varphi^4 \, .
\end{equation}
Under this form, the self-interaction term appears as a local product of $\varphi(x)$, whereas all the information about WNL is recast in the kinetic term\footnote{It is always possible to move the form factor $e^{\vartheta(z)}$ from the kinetic to the interaction terms via field redefinitions, whereas the converse is not true in general.}. The 2 descriptions are strictly equivalent \cite{Briscese:2018oyx}. In the following, only the initial form of the Lagrangian \eqref{L_phi_1} is considered, and the different vacua are discussed. For concreteness, one chooses an SFT-like form factor with $\vartheta(z) = z$, but another choice should give qualitatively similar results.

\paragraph{Klein-Gordon Scalar:}
If $M^2 \geq 0$, there is only 1 vacuum $\phi_0(x) = 0$ that is a minimum of $V(\phi)$, and the field fluctuations around it have the Euclidean propagator
\begin{equation}
\dfrac{-i}{p_E^2 + M^2} \, ,
\end{equation}
that describes only 1 scalar mode $(p_E^2=-M^2)$. The vacuum is tachyon- and ghost-free, and the WNL scale $\Lambda$ acts as a smooth UV-cutoff for the radiative corrections \cite{Namsrai:1986md, Biswas:2014yia, Ghoshal:2017egr, Frasca:2020jbe}. Therefore, the theory is expected to be stable at the perturbative level.

\paragraph{Tachyon Condensation:}
If $M^2 = - \mu^2 < 0$, there is tachyonic instability, and the system relaxes in 1 of the tachyonic vacua, i.e. a (local) minimum of $V(\phi)$, if it exists. This phenomenon is called tachyon condensation.

\subparagraph{\textit{Symmetric Vacuum:}}
The vacuum $\phi_0(x) = 0$ is $\mathbb{Z}_2$-symmetric but unstable, i.e. a local maximum of $V(\phi)$. The fluctuations around this vacuum have the Euclidean propagator
\begin{equation}
\dfrac{-i}{p_E^2 - \mu^2} \, ,
\end{equation}
that is ghost-free and describes only 1 scalar mode: a tachyon (on-shell condition $p_E^2=\mu^2$), reflecting the instability. Therefore, the system must relax in 1 stable vacuum.

\subparagraph{\textit{Broken Vacua:}}
One considers field fluctuations around a constant background, which is the vacuum expectation values (VEV) of the field $\phi(x)$.
There is SSB of the $\mathbb{Z}_2$-symmetry by the choice of 1 of the 2 vacua defined by local minima of $V(\phi)$ that correspond to the VEV's of the field
\begin{equation}
\phi_\pm (x) = \pm v = \pm \sqrt{\dfrac{6}{\lambda}} \, \mu \, ,
\end{equation}
with $\phi_+ \leftrightarrow \phi_-$ under the $\mathbb{Z}_2$-reflection. The field fluctuations $\phi(x) = v +\sigma(x)$ around the $\phi_+$-vacuum\footnote{The physical properties of the fluctuations around the $\phi_-$-vacuum are identical by $\mathbb{Z}_2$-reflection.} are described by the Lagrangian
\begin{equation}
\mathcal{L}_\sigma = -\dfrac{1}{2} \, \sigma \square \sigma + \dfrac{\mu^2}{2} \, \sigma^2 - \dfrac{3\mu^2}{2} \, \varsigma^2 - \dfrac{\mu^2}{v} \, \varsigma^3 - \dfrac{\lambda}{4!} \, \varsigma^4 \, ,
\end{equation}
with the smeared field
\begin{equation}
\varsigma(x) = e^{\frac{1}{2} \, \vartheta \left( - \eta \, \square \right)} \, \sigma(x) \, .
\end{equation}
The Euclidean propagator of the meson field, i.e. the scalar fluctuations $\sigma(x)$ is
\begin{equation}
\dfrac{-i}{p_E^2 - \mu^2 \left[ 1-3 e^{-\eta \, p_E^2} \right]} \, ,
\end{equation}
whose poles $p_E^2 = -m_k^2$ (on-shell condition for a particle of mass $m_k$) are solutions of the transcendental equation
\begin{equation}
e^{-\eta \, m_k^2} \left( m_k^2 + \mu^2 \right) - 3 \mu^2 = 0 \, .
\end{equation}
One introduces the adimensional quantities $\kappa = \mu^2 / \Lambda^2 \geq 0$ and $\varrho_k = m_k^2 / \mu^2$, such that the transcendental equation becomes
\begin{equation}
e^{- \kappa \varrho_k} \left( \varrho_k + 1 \right) - 3 = 0 \, ,
\end{equation}
whose solutions give the meson spectrum
\begin{align}
\varrho_0 &= 2 \, , \ \ \ \text{if} \ \kappa = 0 \, ; \\
\varrho_k &= -1 - \dfrac{W_k \left( -3 \kappa e^{-\kappa} \right)}{\kappa} \, , \ \ \ k \in \mathbb{Z}, \ \ \ \text{if} \ \kappa > 0 \, ,
\end{align}
where $W_k(z)$ is the $k^\text{th}$ branch of the Lambert-$W$ function\footnote{The \textit{Lambert-$W$ function} is also called \textit{omega function} or \textit{product logarithm} in the literature.} \cite{Valluri:2000zz}, defined as the multivalued inverse of the function $f(z) = z e^z$. Let be $\kappa_1  = - W_0(-1/3e) \simeq 0.14$ and $\kappa_2  = - W_{-1}(-1/3e) \simeq 3.29$, there is no real zero when $\kappa \in \left[ \kappa_1, \kappa_2 \right]$, and only 1 or 2 otherwise. When $\kappa > 0$, one gets an infinite tower of complex conjugate zeroes.

\begin{figure}[h!]
\begin{center}
\includegraphics[height=7.7cm]{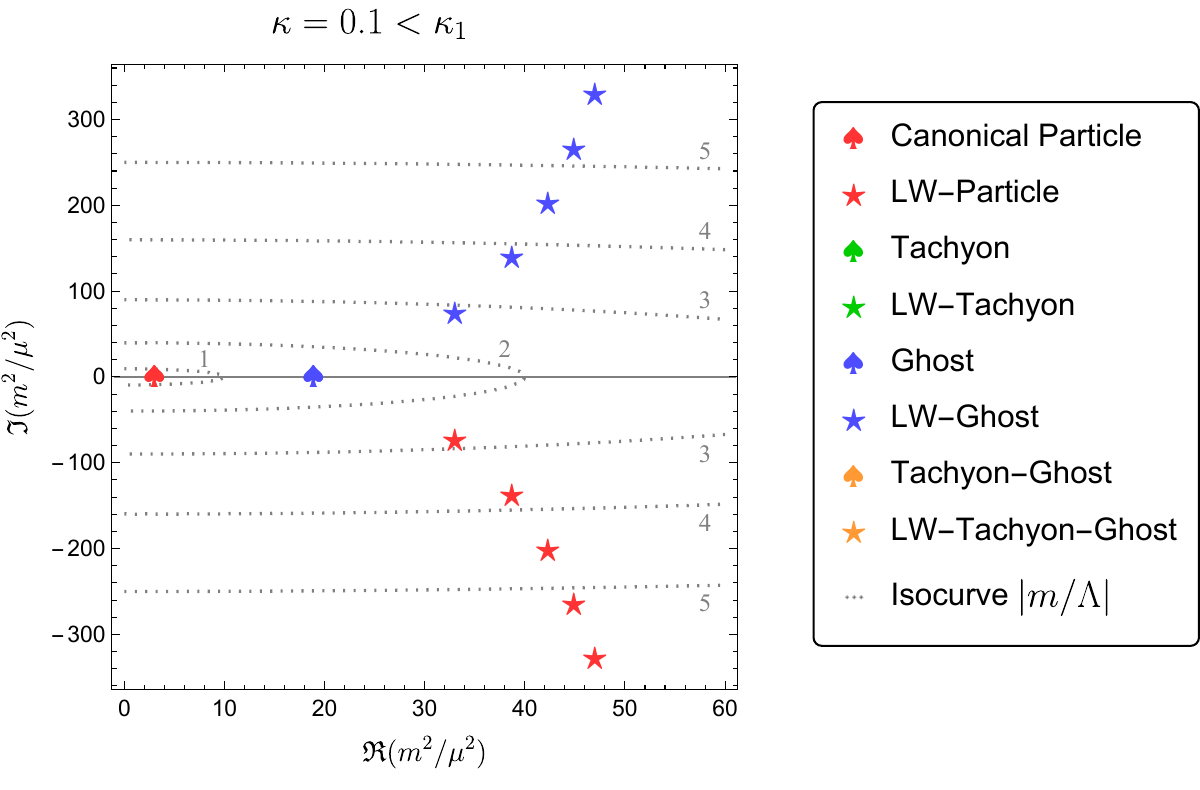}\\
\includegraphics[height=7.7cm]{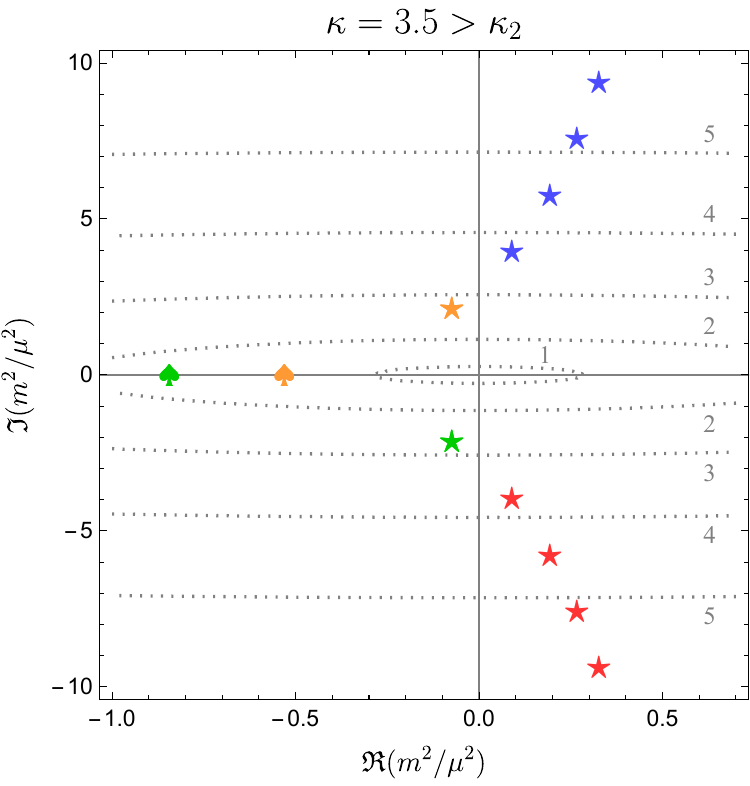}\hfill
\includegraphics[height=7.7cm]{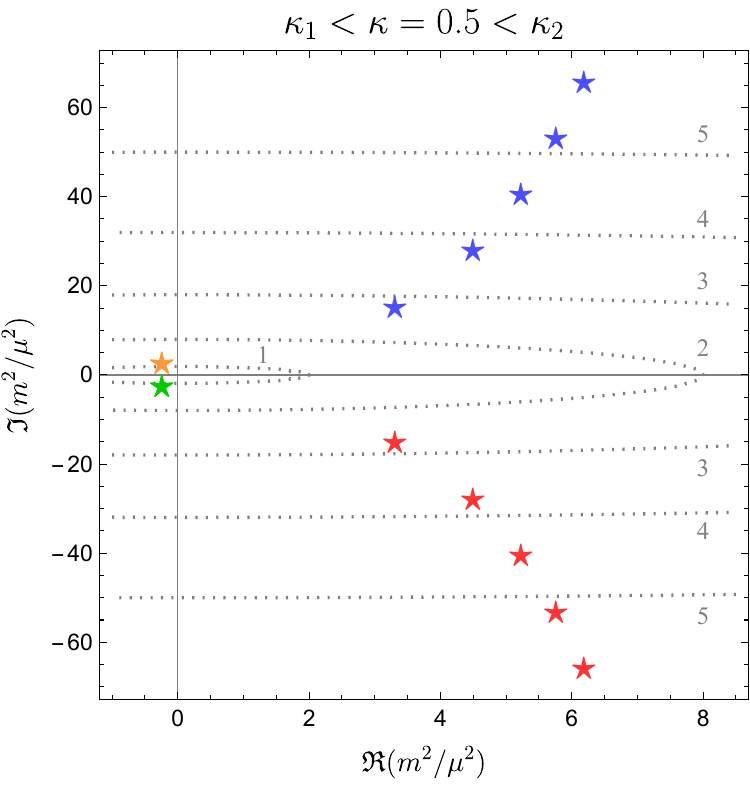}\hfill
\end{center}
\caption{\textit{Meson spectrum in the $(m^2/\mu^2)$-complex plane for the 3 different regimes: normal hierarchy $\mu \ll \Lambda$ (top); inverted hierarchy $\mu \gg \Lambda$ (bottom-left); ahierarchy $\mu \sim \Lambda$ (bottom-right).}}
\label{roots}
\end{figure}

In order to analyze the different physical situations, one can distinguish 3 regimes that lead to qualitatively different meson spectra (cf. Fig~\ref{roots}):
\begin{itemize}[label=$\spadesuit$]
\item \textit{Normal Hierarchy}: When $\mu / \Lambda \ll 1$, one has $\kappa < \kappa_1$, and the meson spectrum is: 1 light canonical mode $(m_0 \sim \mu)$, a ghost around the WNL scale $(m_{-1} \sim \Lambda)$, and an infinite tower of LW-pairs\footnote{LW-pair $=$ 1 LW-particle $+$ 1 LW-ghost, whose masses are the complex conjugates of each other.} triggered by $\Lambda$. In the local limit $(\kappa \rightarrow 0)$, the scale $\Lambda$ decouples and one is left with only the canonical mode $(m_0 = \sqrt{2} \mu)$: one recovers the unique meson mode in local tachyon condensation around the stable vacuum.
\item \textit{Inverted Hierarchy}:  When $\mu / \Lambda \gg 1$, one has $\kappa > \kappa_2$, and the meson spectrum is: 1 tachyon, 1 tachyon-ghost, and an infinite tower of LW-pairs triggered by the scale $\Lambda$.
\item \textit{Ahierarchy}: The case $\kappa \in \left[ \kappa_1, \kappa_2 \right]$ is realized when $\mu / \Lambda \sim 1$, and the meson spectrum is an infinite tower of LW-pairs triggered by $\Lambda$. Real modes appear only when there is a hierarchy between the mass scale in the scalar potential $\mu$ and the WNL scale $\Lambda$.
\end{itemize}

In all regimes, ghosts (and sometimes tachyons) arise in the 2 broken vacua, where stability and/or unitarity are then questioned. The normal hierarchy appears in UV-completions of EFT's in particle physics and/or cosmology, where the ghost-like modes pop up\footnote{Some authors \cite{Barnaby:2007ve, Hashi:2018kag} were puzzled by the interpretation of the spurious ghost-tower in cubic SFT truncated at 0-level, suggesting they might be the precursors of the string modes. However, in a stringy UV-completion, the WNL scale $\Lambda$ corresponds to the string scale, and an infinite tower of string excitations appears. Thus, the truncated SFT-tower they considered cannot be trusted above the string scale \cite{Burgess:2023pnk}: these ghost-like modes are not necessary there. To properly define the EFT, truncating the tower at 0-level is not enough: one should also perform a (local) Taylor expansion $\mu \ll \Lambda$ of the form factor, where only the light mode(s) is (are) propagating in the EFT \cite{Burgess:2023pnk}.} only near and above the WNL scale $\Lambda$. The 2 other regimes do not have light canonical particles; they are plagued with instabilities, and I am not aware of a phenomenological motivation to consider them in model-building.

\subsection{Origin of the Clash \& A Limited Solution}
\label{1st_solution}
The reason behind the reappearance of the ghost problem in $\partial^\infty$QFT with tachyon condensation comes from putting together 2 apparently ``incompatible'' ideas:
\begin{itemize}[label=$\spadesuit$]
\item In string-inspired $\partial^\infty$QFT, the absence of ghosts and tachyons in the particle spectrum is realized by smearing the fields with very specific form factors that do not introduce new poles in the field propagators \cite{Buoninfante:2018mre}. In order to have a true WNL theory (i.e. not just a field redefinition of a local theory), only the fields in the interaction terms are smeared.

\item In tachyon condensation, one expands the scalar field(s) around a (local) minimum of their potential; the mass term of the meson modes (i.e. the fluctuations around the vacuum) comes from both the mass and self-interaction terms of the scalar potential.
\end{itemize}
Therefore, by relaxing in another vacuum via tachyon condensation, one destroys the ghost-free property of the form factor.

A simple solution was proposed in Refs.~\cite{Galli:2010qx, Biswas:2012ka, Koshelev:2020fok}. The idea is to have 2 mass terms: 1 with local fields and 1 with delocalized ones, that conspire together to make the theory ghost-free in the physical vacuum of the theory. Indeed, the symmetric vacuum is already unstable, so there is no big issue with adding ghosts in this vacuum. The core of this idea is to make the theory ghost-free in (at least) 1 global minimum of the potential (it can be degenerate), since it defines the vacuum of the theory from which one defines the asymptotic states of the $S$-matrix. If one takes seriously the ghost problem, the conspiration of the 2 mass terms is not a fine-tuning because the theory needs to protect itself against physical inconsistencies, in the same way as fermion hypercharges must conspire in the SM to cancel gauge anomalies. In order to see how it works, one replaces the previous potential by
\begin{equation}
V (\phi) = -\dfrac{\mu^2}{2} \, \phi^2 + \dfrac{\delta \mu^2}{2} \, \varphi^2 + \dfrac{\lambda}{4!} \, \varphi^4 \, ,
\end{equation}
where the new $\varphi^2$-term looks like a tree-level counter-term with $\mu^2-\delta \mu^2 > 0$ and $\lambda > 0$. One can repeat the analysis of Section~\ref{vacua_tachyons} of the meson fluctuations $\phi(x) = v+\sigma(x)$ around the vacuum
\begin{equation}
v = \sqrt{\dfrac{6}{\lambda} \left( \mu^2 - \delta \mu^2 \right)} \, ,
\end{equation}
and one gets the following Lagrangian for the meson field:
\begin{align}
\mathcal{L}_\sigma
= - \dfrac{1}{2} \, \sigma \square \sigma
+ \dfrac{1}{2} \left( \dfrac{\lambda v^2}{6} + \delta \mu^2 \right) \sigma^2
- \dfrac{1}{2} \left( \dfrac{\lambda v^2}{2} + \delta \mu^2 \right) \varsigma^2 - \dfrac{\lambda v}{3!} \, \varsigma^3 - \dfrac{\lambda}{4!} \, \varsigma^4 \, .
\end{align}
From the discussion in Section~\ref{vacua_tachyons}, the $\varsigma^2$-term must cancel to have a ghost-free meson propagator. This is achieved by adjusting the counter-term coefficient:
\begin{equation}
\delta \mu^2 = - \dfrac{\lambda v^2}{2} \, .
\end{equation}
One can then read the mass of the unique meson mode from the $\sigma^2$-term:
\begin{equation}
m_\sigma = \sqrt{\dfrac{\lambda}{3}} \, v \, ,
\end{equation}
which is identical to the usual relation between the VEV and the quartic coupling in the local limit.

The apparent clash between ghost-freedom and SSB in $\partial^\infty$QFT is not restricted to discrete symmetries; it is also present with both global and gauge continuous symmetries, where the meson field plays the role of the radial fluctuations. In Ref.~\cite{Hashi:2018kag}, the Higgs mechanism is studied in a similar WNL theory, and the physical spectrum also exhibits pathological modes. The problem in this case is that the gauge boson spectrum is also infected by ghosts through its couplings to the Higgs field. For a gauge field, it is not possible to add a gauge-invariant mass counter-term in the manifestly symmetric Lagrangian, as made in the previous example for a scalar field. Therefore, if one wants to realize tachyon condensation (in particular the Higgs mechanism) in an $\partial^\infty$QFT without ghosts and tachyons, one needs a recipe for WNL form factors that adapt themselves to a change of vacuum, i.e. to not lose their good analytic properties when a field gets a VEV. In the rest of this article, an example of such a recipe with a new star-product is proposed: products of fields in the Lagrangian are smeared in both quadratic and quartic terms.

\section{New Weakly Nonlocal Deformation}
\label{WNL_deformation}

\subsection{Star-Product of Fields}
\label{star_product_section}
Let $\Phi_i(x)$ be a set of bosonic fields (fundamental or composite) defined on $\mathbb{R}^{1, 3}$. Spacetime or internal space indices attached to the fields are not written explicitly for clarity. The index $i \in \mathbb{N}$ is understood as labeling the different field species, and the Einstein convention of summation on repeated indices will not be used for it. One defines a \textit{star-product} $\star$ as\footnote{This star-product is formally defined on $\mathbb{R}^{1, 3}$ (Lorentzian signature), but it should be understood as obtained by a formal Wick rotation of the analogous star-product on $\mathbb{R}^4$ (Euclidean signature). A reader familiar with noncommutative QFT should find some similarity with the Groenewold–Moyal star-product, which is noncommutative and breaks explicitly Lorentz invariance, cf. Refs~\cite{Douglas:2001ba, Szabo:2001kg} for reviews. Instead, the star-product considered in this article will be designed to be commutative and Lorentz covariant.} (no summation on $i$ and $j$)
\begin{align}
\Phi_i(x) \star \Phi_j(x)
&= \Phi_i(x) \, e^{\vartheta_{ij} \left( \overleftarrow{\partial_\mu} \, \eta^{\mu \nu}_{ij} \, \overrightarrow{\partial_\nu}  \right)} \, \Phi_j(x) \, , \nonumber \\
&= \left. e^{\vartheta_{ij} \left( \partial_\mu^{(i)} \, \eta^{\mu \nu}_{ij} \, \partial_\nu^{(j)} \right)} \, \Phi_i(x_i) \cdot \Phi_j(x_j) \right|_{x_i \to x_j \equiv x} \, , \nonumber \\
&= \int \dfrac{d^4p_i \, d^4p_j}{(2 \pi)^8} \ e^{-i \left( p_i + p_j \right) \cdot x \, + \, \vartheta_{ij} \left( -\eta_{ij} \, p_i \cdot p_j \right)} \, \widetilde{\Phi}_i(p_i) \cdot \widetilde{\Phi}_j(p_j) \, ,
\label{star-def}
\end{align}
with the notation $\eta^{\mu \nu}_{ij} = \eta_{ij} \, g^{\mu \nu}$, where $\eta_{ij} \in \mathbb{C}$ is a scalar whose modulus is homogeneous to an area in natural units. Note that the star-product depends on the 2 involved field species via both $\eta_{ij}$ and the entire function $\vartheta_{ij}(z \in \mathbb{C})$, such that
\begin{equation}
\forall z \in \mathbb{C} , \ 
e^{\vartheta_{ij}(z)} = \sum_{n=0}^{+\infty} c_{ij}^{(n)} \, z^n \, ,
\end{equation}
satisfying $\vartheta_{ij}(0)=0$, thus (no summation on $i$ and $j$)
\begin{align}
\Phi_i(x) \star \Phi_j(x)
= \Phi_i(x) \cdot \Phi_j(x)
&+ c_{ij}^{(1)} \eta_{ij} \, \partial_\mu \Phi_i(x) \cdot \partial^\mu \Phi_j(x) \nonumber \\
&+ c_{ij}^{(2)} \eta_{ij}^2 \, \partial_\mu \partial_\nu \Phi_i(x) \cdot \partial^\mu \partial^\nu \Phi_j(x)
+ \mathcal{O} \left( \eta_{ij}^3 \right) .
\end{align}
The star-product is a WNL deformation of the usual (local) pointwise product $\cdot$ since, in the local limit, one has
\begin{equation}
\Phi_i(x) \star \Phi_j(x) \underset{\eta_{ij} \to 0}{\sim} \Phi_i(x) \cdot \Phi_j(x) \, .
\end{equation}
If a field is equal to its VEV: $\Phi_i(x) = v_i$ then, by construction:
\begin{equation}
v_i \star \Phi_j(x) = v_i \cdot \Phi_j(x) \, .
\label{vev_prod}
\end{equation}

One can impose further restrictions on this multiplicative law $\star$ to preserve as much as possible the properties of the pointwise product in local QFT:
\begin{itemize}[label=$\spadesuit$]
\item Commutativity:
\begin{align}
\Phi_i(x) \star \Phi_j(x)
= \Phi_j(x) \star \Phi_i(x) \, ,
\end{align}
with $\forall z \in \mathbb{R} , \ \vartheta_{ij}(z) = \vartheta_{ji}(z)$ and $\eta_{ij} = \eta_{ji}$.
\item Hermitian symmetry:
\begin{align}
\Phi_i(x) \star \Phi_j(x)
= \left[ \Phi_j(x) \star \Phi_i(x) \right]^* \, ,
\end{align}
with $\forall z \in \mathbb{R} , \ \vartheta_{ij}(z) = \vartheta_{ji}(z)^*$ and $\eta_{ij} = \eta_{ji}^*$. Together with commutativity, one gets that the $\eta_{ij}$'s and $\vartheta_{ij}(z \in \mathbb{R})$'s are real-valued symmetric matrices.
\item Linearity in both arguments:
\begin{align}
&\left[ \lambda_i \, \Phi_i(x) + \lambda_j \, \Phi_j(x) \right]
\star \left[ \lambda_k \, \Phi_k(x) + \lambda_l \, \Phi_l(x) \right] \nonumber \\
&= \lambda_i \, \lambda_k \, \Phi_i(x) \star \Phi_k(x)
+ \lambda_i \, \lambda_l \, \Phi_i(x) \star \Phi_l(x) \nonumber \\
&+ \lambda_j \, \lambda_k \, \Phi_j(x) \star \Phi_k(x)
+ \lambda_j \, \lambda_l \, \Phi_j(x) \star \Phi_l(x) \, ,
\end{align}
where the usual additive law $+$ is defined between the fields with the same quantum numbers (i.e. in the same representation of the symmetry group or its complex conjugate). It imposes that $\forall z \in \mathbb{R} , \ \vartheta_{ik}(z) = \vartheta_{jk}(z) = \vartheta_{il}(z) = \vartheta_{jl}(z)$ and $\eta_{ik} = \eta_{jk} = \eta_{il} = \eta_{jl}$. With the previous properties, this means that for each couple $(a,b)$ of group representations, one has a different form factor exponent $\vartheta_{ab}(z)$ and a different fuzzy plaquette $\eta_{ab}$.
\end{itemize}
However, by an analysis in momentum space, one can check easily that this star-product is not associative in general\footnote{A notable exception with associativity is when $\forall (i, j)$, $\eta_{ij} \equiv \eta$ and $\vartheta_{ij}(z) \equiv \vartheta(z) \propto z$, i.e. SFT-like.}, i.e.
\begin{align}
\left[ \Phi_i(x) \star \Phi_j(x) \right] \star \Phi_k(x)
\neq \Phi_i(x) \star \left[ \Phi_j(x) \star \Phi_k(x) \right] \, .
\label{assoc}
\end{align}

In order to apply this framework in QFT, one can replace the pointwise products in the local operator products with this new star-product. Such WNL deformation of the original local theory defines a whole class of possible $\partial^\infty$QFT's, without further selection criteria\footnote{It is not so different from the ad hoc prescriptions to define other string-inspired QFT's \cite{Buoninfante:2018mre, BasiBeneito:2022wux}.}. Indeed, the different $\vartheta_{ij}(z)$'s and $\eta_{ij}$'s allow large freedom but can be at least constrained by the reasonable requirement to not break an exact symmetry of the original local theory. Moreover, since the star-product is not associative in general, a local operator product like $\Phi_i(x) \cdot \Phi_j(x) \cdot \Phi_k(x)$ can be deformed as different associations of star-products between the same fields, cf. Eq.~\eqref{assoc}.

\subsection{Vacuum Stability in Tachyon Condensation}
\label{star_fuzzy_tachyon_scalar}
In this section, the star-product is applied to get an alternative WNL deformation of the toy scalar model of Section~\ref{vacua_tachyons}. Since there is only 1 real scalar field $\phi(x)$, one needs to introduce only 1 fuzzy plaquette $\eta$ and 1 entire function $\vartheta(z)$. Consider the Lagrangian
\begin{equation}
\mathcal{L}_\phi^{(\star)} = \dfrac{1}{2} \, \partial^\mu \phi \star \partial_\mu \phi - V^{(\star)}(\phi) \, ,
\label{L_phi_2}
\end{equation}
with the potential $(M^2 \in \mathbb{R}, \ \lambda_{1, 2} > 0)$:
\begin{equation}
V^{(\star)}(\phi) = \dfrac{M^2}{2} \, \phi \star \phi
+ \dfrac{\lambda_1}{4!} \left( \phi \star \phi \right) \star \left( \phi \star \phi \right) + \dfrac{\lambda_2}{4!} \, \phi \star \left[ \phi \star \left( \phi \star \phi \right) \right] \, ,
\end{equation}
where one has included the 2 nonequivalent quartic self-interaction terms to illustrate the nonassociativity of the star-product. It reduces to the Lagrangian \eqref{L_phi_1} with the potential \eqref{V_phi_1} in the local limit $\eta \rightarrow 0$, where $\lambda_1 + \lambda_2 = \lambda$. The WNL Lagrangian is still manifestly invariant under the $\mathbb{Z}_2$-reflection $\phi(x) \mapsto - \phi(x)$.

\paragraph{Klein-Gordon Scalar:}
In the case $M^2 \geq 0$, there is still only 1 vacuum $\phi_0(x) = 0$ that is $\mathbb{Z}_2$-symmetric. One performs the field redefinition \eqref{field_redef} that leaves the vacuum and $S$-matrix identical. It is instructive to analyze the action in momentum space.

\subparagraph{\textit{Quadratic Terms:}} The mass term involves the field product
\begin{align}
\int d^4x \ \phi(x) \star \phi(x) &= \int \dfrac{d^4p}{(2 \pi)^4} \ e^{\vartheta(\eta \, p^2)} \, \widetilde{\phi}(p) \cdot \widetilde{\phi}(-p) \, , \\
&= \int \dfrac{d^4p}{(2 \pi)^4} \ \widetilde{\varphi}(p) \cdot \widetilde{\varphi}(-p) \, , \\
&= \int d^4x \ \varphi^2(x) \, .
\end{align}
In the same way, the kinetic term gives
\begin{align}
\int d^4x \ \partial^\mu \phi(x) \star \partial_\mu \phi(x) = \int d^4x \ \partial^\mu \varphi(x) \cdot \partial_\mu \varphi(x) \, .
\end{align}
Therefore, the field redefinition \eqref{field_redef} removes the WNL features from the quadratic terms. The Euclidean propagator
\begin{equation}
\dfrac{-i}{p_E^2+M^2}
\end{equation}
describes only 1 propagating scalar mode $(p_E^2 = -M^2)$, and the vacuum is thus tachyon- and ghost-free.

\subparagraph{\textit{Quartic Terms:}} The $1^\text{st}$ self-interacting term is
\begin{align}
\int d^4x \left( \phi \star \phi \right) \star \left( \phi \star \phi \right)
= \int \left[ \prod_{i=1}^4 \dfrac{d^4 p_i}{(2 \pi)^4} \ \widetilde{\varphi} \left( p_i \right) \right] (2 \pi)^4 \, \delta \left( \sum_{i=1}^4 p_i \right) \, e^{\vartheta_1^{(4)}(p_1, \, p_2, \, p_3, \, p_4)} \, ,
\end{align}
with
\begin{align}
\vartheta_1^{(4)}(p_1, p_2, p_3, p_4)
&= \vartheta \left(- \eta \, p_1 \cdot p_2 \right) + \vartheta \left(- \eta \, p_3 \cdot p_4 \right) + \vartheta \left[- \eta \left( p_1 + p_2 \right) \cdot \left( p_3 + p_4 \right) \right] \nonumber \\
&- \dfrac{1}{2} \sum_{i=1}^4 \vartheta \left( \eta \, p_i^2 \right) \, ,
\end{align}
and the $2^\text{nd}$ one is
\begin{align}
\int d^4x \ \phi \star \left[ \phi \star \left( \phi \star \phi \right) \right] = \int \left[ \prod_{i=1}^4 \dfrac{d^4 p_i}{(2 \pi)^4} \ \widetilde{\varphi} \left( p_i \right) \right] (2 \pi)^4 \, \delta \left( \sum_{i=1}^4 p_i \right) \, e^{\vartheta_2^{(4)}(p_1, \, p_2, \, p_3, \, p_4)} \, ,
\end{align}
with
\begin{align}
\vartheta_2^{(4)}(p_1, p_2, p_3, p_4)
&= \vartheta \left(- \eta \, p_1 \cdot p_2 \right) + \vartheta \left[- \eta \, p_3 \cdot \left( p_1 + p_2 \right) \right] + \vartheta \left[- \eta \, p_4 \cdot \left( p_1 + p_2 + p_3 \right) \right] \nonumber \\
&- \dfrac{1}{2} \sum_{i=1}^4 \vartheta \left( \eta \, p_i^2 \right) \, ,
\end{align}
where the nonassociativity of the star-product is now manifest for a general\footnote{In this toy model, one can check that associativity holds $(\vartheta_1^{(4)} = \vartheta_2^{(4)})$ for $\vartheta(z) \propto z$.} $\vartheta(z)$. The WNL features are recast in interaction terms. From their analytic structures, the associated vertices do not introduce any new singularity (pole or branch cut) at tree-level.

\paragraph{Tachyon Condensation:}
In the case of the tachyonic instability $M^2 = - \mu^2<0$, there is 1 $\mathbb{Z}_2$-symmetric vacuum $\phi_0=0$ that is a local maximum of $V^{(\star)}(\phi)$, and 2 degenerate vacua
\begin{equation}
\phi_\pm = \pm v = \pm \sqrt{\dfrac{6}{\lambda_1+\lambda_2}} \, \mu \, , 
\end{equation}
that are local minima of $V^{(\star)}(\phi)$ with SSB of the $\mathbb{Z}_2$ symmetry. The mesonic fluctuations $\phi(x) = v + \sigma(x)$ around the $\phi_+$-vacuum have the Lagrangian
\begin{align}
\mathcal{L}_\sigma^{(\star)}
&=  \dfrac{1}{2} \, \partial^\mu \sigma \star \partial_\mu \sigma - \mu^2 \, \sigma \star \sigma - \dfrac{\mu^2}{v} \, \sigma \star \left( \sigma \star \sigma \right) \nonumber \\
&- \dfrac{\lambda_1}{4!} \left( \sigma \star \sigma \right) \star \left( \sigma \star \sigma \right) - \dfrac{\lambda_2}{4!} \, \sigma \star \left[ \sigma \star \left( \sigma \star \sigma \right) \right] \, .
\end{align}
A similar analysis to the Klein-Gordon scalar shows that there is only 1 perturbative degree of freedom in all vacua: 1 tachyon of mass $i \mu$ in the $\phi_0$-vacuum, and 1 meson of mass $\sqrt{2} \mu$ in the $\phi_\pm$-vacua, such as in the local limit $\eta \rightarrow 0$. This is transparent in the form of the Lagrangian, where the kinetic and interaction terms are equally treated in this WNL formalism. Note that the meson mass does not receive contributions from the WNL scale $\Lambda = \sqrt{1/\eta}$, contrary to the smeared field formalism of Section~\ref{vacua_tachyons}. One concludes that the local minima $\phi_\pm$ are stable at tree-level.

\subsection{Application to Gauge Symmetry Breaking}
One of the most beautiful applications of tachyon condensation is the SSB of gauge symmetries by the Higgs mechanism \cite{Nambu:1960tm, Anderson:1963pc, Higgs:1964ia, Englert:1964et, Higgs:1964pj, Guralnik:1964eu, Higgs:1966ev, Migdal:1966tq, Kibble:1967sv, Guralnik:2011zz}. In order to keep local gauge transformations and to generalize the star-product in a manifestly gauge covariant way, one uses the Krasnikov-Terning scheme\footnote{The Krasnikov-Terning scheme is sharply different from the deformed gauge transformations with a star-product in noncommutative QFT's that are motivated by quantum spacetimes \cite{Hersent:2022gry}, e.g. the Seiberg-Witten map \cite{Seiberg:1999vs} in string theory.} \cite{Krasnikov:1987yj, Terning:1991yt, Biswas:2011ar}, i.e. any partial derivative is replaced by a covariant derivative $(\partial_\mu \mapsto \mathcal{D}_\mu)$ in the star-product definition \eqref{star-def}. In the following, one will illustrate this with the example of a WNL deformation of an Abelian Higgs mechanism. It is important to stress that, in the proposed formalism, both pointwise and star-products can coexist\footnote{The coexistence of the 2 products is again different from noncommutative QFT's \cite{Douglas:2001ba, Szabo:2001kg}, based on a deformation of the original algebra of local QFT, where all pointwise products are replaced by star-products, even in the series expansion of entire functions.}. In this article, all star-products are written explicitly by using the symbol $\star$, the other explicit/implicit products have to be understood as pointwise.

One will discuss the case of an Abelian Higgs mechanism with a toy model that gives mass to a fictitious photon. Consider a photon field $A_\mu(x)$ and a Higgs field $\phi(x)$ with an electric charge $-1$. The electromagnetic gauge coupling is $e$ and is associated with the gauge group $U(1)_{em}$. One wants to keep local gauge invariance, so the field transformations under $U(1)_{em}$ are the standard ones
\begin{equation}
\phi(x) \mapsto e^{i e \, \alpha(x)} \, \phi(x) \, , \ \ \ 
A_\mu(x) \mapsto A_\mu(x) - \partial_\mu \alpha(x) \, ,
\label{gauge_transfo}
\end{equation}
where the exponential function is defined via its power series built with the (local) pointwise product. In order to avoid any confusion, one will define 2 different symbols for the star-product between fields with different quantum numbers.

For neutral fields, one introduces the fuzzy plaquette $\eta_0 = 1/\Lambda_0^2$ and the symbol $\star_0$ defined as (no summation on $i$ and $j$)
\begin{align}
A_{\rho} (x) \star_0 A_{\sigma} (x)
&= A_{\rho}(x) \, e^{\vartheta_0 \left( \overleftarrow{\partial_\mu} \, \eta^{\mu \nu}_0 \, \overrightarrow{\partial_\nu} \right)} \, A_{\sigma}(x) \, , \\
&= \left. e^{\vartheta_0 \left( \partial_\mu^{(i)} \, \eta^{\mu \nu}_0 \, \partial_\nu^{(j)} \right)} \, A_{\rho}(x_i) \cdot A_{\sigma}(x_j) \right|_{x_i \to x_j \equiv x} \, ,
\end{align}
with the entire function $\vartheta_0(z)$ and the notation $\eta^{\mu \nu}_0 = \eta_0 \, g^{\mu \nu}$.

For fields creating/annihilating particles of charge $-1$ and antiparticles of charge $+1$, one introduces the fuzzy plaquette $\eta_e = 1/\Lambda_e^2$ and the symbol $\star_e$ defined as (no summation on $i$ and $j$)
\begin{align}
\phi^*(x) \star_e \phi(x)
&= \phi^*(x) \, e^{\vartheta_e \left( \overleftarrow{\mathcal{D}_\mu} \, \eta^{\mu \nu}_e \,\overrightarrow{\mathcal{D}_\nu} \right)} \, \phi(x) \, , \label{star-e00} \\
&= \left. e^{\vartheta_e \left( \overline{\mathcal{D}}_\mu^{(i)} \, \eta^{\mu \nu}_e \, \mathcal{D}_\nu^{(j)} \right)} \, \phi^*(x_i) \cdot \phi(x_j) \right|_{x_i \to x_j \equiv x} \, , \label{star-e0} \\
&= \left[ \int \dfrac{d^4 p_i \, d^4 p_j}{(2 \pi)^8} \, e^{-i \left( p_i + p_j \right) \cdot x} \, \widetilde{\phi^*}(p_i) \cdot \phi(p_j) \right] \left[ \left. e^{\vartheta_e \left( -\eta_e \, P_{ij} \right)} \right|_{x_i \to x_j \equiv x} \right] \, ,
\label{star-e}
\end{align}
with the entire function $\vartheta_e(z)$ and the notations: $\eta^{\mu \nu}_e = \eta_e \, g^{\mu \nu}$;
\begin{equation}
P_{ij} = p_i \cdot p_j + i p_j \cdot \overline{\mathcal{D}}^{(i)} + i p_i \cdot \mathcal{D}^{(j)} - \overline{\mathcal{D}}^{(i)} \cdot \mathcal{D}^{(j)} \, .
\end{equation}
The covariant derivatives are defined as
\begin{equation}
\mathcal{D}_\mu^{(i)} = \partial_\mu^{(i)} + i e \, A_\mu(x_i) \, ,
 \ \ \ 
\overline{\mathcal{D}}_\mu^{(i)} = \partial_\mu^{(i)} - i e \, A_\mu(x_i) \, ,
\end{equation}
where $\left[ \overline{\mathcal{D}}_\mu^{(i)} , \mathcal{D}_\nu^{(j)} \right] = 0$ with $x_i \neq x_j$. To get Eq.~\eqref{star-e} from Eq.~\eqref{star-e0}, one uses the identity
\begin{equation}
f ( \partial ) \, e^{-ip \cdot x} = e^{-ip \cdot x} \, f ( \partial - i p ) \, .
\end{equation}
Note that the action of a covariant derivative on a field uses the pointwise product between the gauge connection and the charged field:
\begin{equation}
\mathcal{D}_\mu^{(i)} \phi(x_i) = \left[ \partial_\mu^{(i)} + i e \, A_\mu(x_i) \right] \cdot \phi(x_i) \, .
\end{equation}
Therefore, one has
\begin{align}
\phi^*(x) \star_e \phi(x)
= \phi^*(x) \cdot \phi(x)
&+ c_e^{(1)} \eta_e \, \overline{\mathcal{D}}_\mu \phi^*(x) \cdot \mathcal{D}^\mu \phi(x) \nonumber \\
&+ c_e^{(2)} \eta_e^2 \, \overline{\mathcal{D}}_\mu \overline{\mathcal{D}}_\nu \phi^*(x) \cdot \mathcal{D}^\mu \mathcal{D}^\nu \phi(x)
+ \mathcal{O} \left( \eta_e^3 \right) .
\end{align}

The WNL Lagrangian for the Abelian Higgs model is
\begin{equation}
\mathcal{L}_\star = - \dfrac{1}{4} \, F_{\mu\nu} \star_0 F^{\mu\nu} + \left( \overline{\mathcal{D}}_\mu \phi^* \right) \star_e \left( \mathcal{D}^\mu \phi \right) - V_\star(\phi) \, ,
\end{equation}
with the Faraday tensor $F_{\mu\nu} = \partial_\mu A_\nu - \partial_\nu A_\mu$ and the scalar potential $V_\star (\phi)$. As usual, one wants to study the fluctuations of the Higgs field around its VEV $v$. One can thus use the polar representation for the Higgs field:
\begin{equation}
\phi(x) = \dfrac{1}{\sqrt{2}} \left[ v + \sigma(x) \right] e^{i\frac{\pi(x)}{v}} \, ,
\end{equation}
with the real scalar fields $\sigma(x)$ and $\pi(x)$ for the meson and pion, respectively. From the discussions in Sections~\ref{star_product_section} and \ref{star_fuzzy_tachyon_scalar}, the most natural attempt for $V_\star(\phi)$ is to replace all pointwise products in the usual local quartic potential of the Higgs field by star-products $(\mu^2 > 0, \ \lambda > 0)$:
\begin{equation}
V_\star(\phi) = - \mu^2 \, \phi^* \star_e \phi + \lambda \left( \phi^* \star_e \phi \right) \star_0 \left( \phi^* \star_e \phi \right) \, .
\end{equation}
In order to be fully explicit, the full expressions of each Lagrangian term with the star-products are
\begin{align}
F_{\mu\nu} \star_0 F^{\mu\nu} &= F_{\rho\sigma} \, e^{\vartheta_0 \left( \overleftarrow{\partial_\mu} \, \eta^{\mu \nu}_0 \, \overrightarrow{\partial_\nu} \right)} \, F^{\rho\sigma} \, , \\
\left( \overline{\mathcal{D}}_\mu \phi^* \right) \star_e \left( \mathcal{D}^\mu \phi \right) &= \overline{\mathcal{D}}_\rho \phi^* \, e^{\vartheta_e \left( \overleftarrow{\mathcal{D}_\mu} \, \eta^{\mu \nu}_e \,\overrightarrow{\mathcal{D}_\nu} \right)} \, \mathcal{D}^\rho \phi \, , \\
\left( \phi^* \star_e \phi \right) \star_0 \left( \phi^* \star_e \phi \right) &= \left( \phi^* \star_e \phi \right) e^{\vartheta_0 \left( \overleftarrow{\partial_\mu} \, \eta^{\mu \nu}_0 \, \overrightarrow{\partial_\nu} \right)} \left( \phi^* \star_e \phi \right) \, .
\end{align}
One can check that the Lagrangian $\mathcal{L}_\star$ is Hermitian symmetric, as well as manifestly Lorentz and gauge invariant. In the local limit $(\eta_{0, e} \rightarrow 0)$, one recovers the (local) pointwise products between fields.

Nevertheless, there is a major issue for this choice of $V_\star (\phi)$. Indeed, the dependence on covariant derivatives in the expression of the star-product between Higgs fields \eqref{star-e00} means that $\phi(x)$ is dressed by a ``photon cloud''. In perturbation theory $(e \ll 1)$, one has $\eta_e \gg e^2 \, \eta_e$ so it is justified to partially expand the pseudodifferential operator in the definition \eqref{star-e00} of the star-product $\star_e$ as
\begin{equation}
e^{\vartheta_e \left( \overleftarrow{\mathcal{D}_\mu} \, \eta^{\mu \nu}_e \,\overrightarrow{\mathcal{D}_\nu} \right)} = e^{\vartheta_e \left( \overleftarrow{\partial_\mu} \, \eta^{\mu \nu}_e \, \overrightarrow{\partial_\nu} \right)} + \mathcal{O}\left( \dfrac{e}{\Lambda_e^2} \, A \right) \, ,
\label{exp}
\end{equation}
in the energy range $(\Lambda_e, \, \Lambda_e/\sqrt{e})$. In this semiperturbative expansion, one performs a perturbative expansion in the gauge coupling while keeping the full momentum dependence of the form factor (nonperturbative). In the ellipsis $\mathcal{O}\left( A \right)$, there is an infinite tower of local operators that give vertices with a growing number of photons and partial derivatives. Therefore, the star-product between Higgs fields $\phi^* \star_e \phi$ includes
\begin{equation}
v \star_e v = v^2 \left. e^{\vartheta_e \left( \overline{\mathcal{D}}_\mu^{(i)} \, \eta^{\mu \nu}_e \, \mathcal{D}_\nu^{(j)} \right)} \right|_{x_i \to x_j \equiv x} = v^2 + \mathcal{O} \left( A^2 \right) \, .
\end{equation}
Again, the ellipsis $\mathcal{O} \left( A^2 \right)$ contains an infinite tower of operators with an increasing number of photon fields and partial derivatives. The problem is that they contribute to the part of the Lagrangian that is quadratic in the photon field $A_\mu(x)$, thus destroying the ghost-free property of the theory\footnote{This major issue was not taken into account in the published version of this article (I realized it later by myself). The present e-print version shows how to avoid it.}.

A simple solution is to use a scalar potential without terms involving $\star_e$, but only $\star_0$ and/or the pointwise product. The general quartic scalar potential fulfilling this requirement is\footnote{When no parentheses are displayed, exponentiation $\phi^n$ or $|\phi|^{2n}$ (defined with the pointwise product) takes priority over star-product.}
\begin{align}
V_\star(\phi)
&= - \mu^2 \, |\phi|^2 + \lambda \, | \phi |^2 \star_0 | \phi |^2 - \delta \lambda \, | \phi |^4 \, ,
\label{final_Higgs_potential}
\end{align}
with $\mu^2 > 0$ and $\left( \lambda - \delta \lambda \right) > 0$. A local quartic counter-term has been included (in the same spirit as in Section~\ref{1st_solution}), since it is not clear (a priori) that the model will be automatically ghost-free in the broken vacuum with a scalar potential involving also pointwise products.

The scalar potential $V_\star(\phi)$ has a continuous set of minima, defined as any $U(1)_{em}$ rotation of the VEV
\begin{equation}
v = \sqrt{\dfrac{\mu^2}{\lambda - \delta \lambda}} \, .
\end{equation}
One works in unitary gauge, i.e. one uses the freedom given by the gauge transformations \eqref{gauge_transfo} to remove the explicit dependence on the pion field $\pi(x)$, which becomes the longitudinal polarization of the massive photon field $A_\mu(x)$, i.e.
\begin{equation}
H(x) = e^{-i \frac{\pi(x)}{v}} \, \phi(x) = \dfrac{1}{\sqrt{2}} \left[ v + \sigma(x) \right] \, ,
\end{equation}
that is now real-valued. One gets the Lagrangian (in unitary gauge):
\begin{align}
\mathcal{L}_\star^U &= - \dfrac{1}{4} \, F_{\mu\nu} \star_0 F^{\mu\nu} + \partial_\mu H \star_e \partial^\mu H  \nonumber \\
&+ e^2 \left( H \cdot A_\mu \right) \star_e \left( H \cdot A^\mu \right) + \mu^2 \, H^2 \nonumber \\
&- \lambda \, H^2 \star_0 H^2 + \delta \lambda \, H^4 \, .
\label{U_Lag}
\end{align}

As discussed previously, the model is designed in such a way that the photon clouds (of the star-products $\star_e$) do not contribute to the quadratic terms of the Lagrangian at tree-level. Therefore, to study the perturbative spectrum of the model, one can freely drop the gauge clouds from the form factor expansion \eqref{exp}. One will note this noncovariant truncation of the star-product with the symbol $\ast_e$, instead of $\star_e$. Concerning the star-product between neutral fields, the partial derivatives are already covariant, so the symbols $\star_0$ and $\ast_0$ are equivalent. Therefore, one has
\begin{equation}
\phi^* \ast_e \phi = \phi^* \, e^{\vartheta_e \left( \overleftarrow{\partial_\mu} \, \eta^{\mu \nu}_e \, \overrightarrow{\partial_\nu} \right)} \, \phi \, ,
\ \ \ 
F_{\mu\nu} \ast_0 F^{\mu\nu} = F_{\mu\nu} \star_0 F^{\mu\nu} \, .
\end{equation}
Then, it is important to notice that, in the broken vacuum, the mass terms for the photon and meson fields receive contributions from the interaction terms in the Lagrangian \eqref{U_Lag}. Therefore, the ghost-free condition is realized if one imposes $\eta \equiv \eta_0 = \eta_e$ and $\vartheta(z) \equiv \vartheta_0(z) = \vartheta_e(z)$, thus one can also drop the indices attached to the symbols $\ast$. At the end, the Lagrangian $\mathcal{L}_\star^U$ can be written in terms of meson and Proca fields as
\begin{align}
\mathcal{L}_\star^U &\supset - \dfrac{1}{4} \, F_{\mu\nu} \ast F^{\mu\nu}
+ \dfrac{1}{2} \, \partial_\mu \sigma \ast \partial^\mu \sigma
+ \dfrac{(e v)^2}{2} \, A_\mu \ast A^\mu
\nonumber \\
&- \lambda v^2 \, \sigma \ast \sigma
- \lambda v \, \sigma \ast \sigma^2
- \dfrac{\lambda}{4} \, \sigma^2 \ast \sigma^2
\nonumber \\
&+ \delta \lambda v^2 \, \sigma^2
+ \delta \lambda v \, \sigma^3
+ \dfrac{\delta \lambda}{4} \, \sigma^4 \, .
\end{align}
The counter-term coefficient $\delta \lambda$ is fixed to cancel the $\sigma^2$-term:
\begin{equation}
\delta \lambda = 0 \, ,
\end{equation}
which means that there is no local quartic counter-term at tree-level in the potential \eqref{final_Higgs_potential}. Then, one performs the field redefinitions
\begin{equation}
A_\mu(x) \mapsto e^{-\frac{1}{2} \, \vartheta \left( -\eta \, \square \right)} \, A_\mu(x) \, , \ \ \ 
H(x) \mapsto e^{-\frac{1}{2} \, \vartheta \left( -\eta \, \square \right)} \, H(x) \, ,
\end{equation}
to recast the pseudodifferential operators in the interaction terms. One concludes that the model describes only 2 physical particles: 1 massive photon of mass $m_A$, and 1 meson (i.e. the Higgs boson!) of mass $m_\sigma$, given by
\begin{equation}
m_A = ev \ \ \ \text{and} \ \ \ m_\sigma = \sqrt{2 \lambda} v \, ,
\end{equation}
with the same relation between the VEV and the couplings as in the local Abelian Higgs model. There is no spurious tachyon or ghost mode to spoil the perturbative stability/unitarity in the physical vacuum.

\section{Conclusion \& Outlook}
\label{conclusion}
In this article, one has tackled the problem of finding a ghost-free $\partial^\infty$QFT with tachyon condensation. In Section~\ref{section_problem}, the case of an SFT-inspired scalar theory with the SSB of a $\mathbb{Z}_2$ symmetry is studied. For a normal hierarchy, an infinite tower of ghost-like particles appears close to the WNL scale at tree-level, which questions the validity of the theory. In Section~\ref{WNL_deformation}, a new WNL deformation is proposed with the definition of a star-product between fields, which is manifestly Lorentz and gauge covariant, commutative, but nonassociative in general. One has shown that no spurious tower pops up in this case. In particular, the example of a ghost-free WNL Abelian Higgs mechanism at the perturbative level appears to be a very promising result for future investigations.

In this study, a solution to the issue of potential pathologies in the perturbative spectrum of simple $\partial^\infty$QFT's has been proposed. In order to further test the validity of such $\partial^\infty$QFT, several aspects need to be investigated, e.g.
\begin{itemize}[label=$\spadesuit$]
\item A generalization of the star-product that includes fermions, non-Abelian gauge symmetries, and gravity should not be the major difficulty.
\item The UV behavior of radiative corrections is particularly interesting, since the goal is to achieve superrenormalizability or UV-finiteness. In order to be under perturbative control with the Krasnikov-Terning scheme \cite{Krasnikov:1987yj, Terning:1991yt, Biswas:2011ar}, it is (a priori) necessary to choose entire functions $\vartheta_{ij}(z)$ that lead to KT form factors \cite{Kuzmin:1989sp, Tomboulis:1997gg, Modesto:2011kw, Modesto:2014lga, Modesto:2015lna, Modesto:2015foa, Tomboulis:2015esa, Giaccari:2016kzy, Modesto:2016max, Modesto:2017hzl, Modesto:2017sdr, Koshelev:2017ebj, BasiBeneito:2022wux}, otherwise one runs into trouble \cite{Talaganis:2014ida}. The reason is that the power counting theorem explicitly assumes UV-locality \cite{Weinberg:1959nj, Collins:1984xc, Anselmi:2019pdm}. However, a power counting in the deep-UV is not sufficient. One also needs to prove the locality of the counter-terms when one probes the WNL regime (i.e. around the nonlocal scale), which is highly nontrivial: a nonlocal form factor depending only on the external momenta may factorize in front of a UV-divergent integral, spoiling the assumptions of the power counting theorem, even with a KT form factor. Unfortunately, this question was only investigated by Tomboulis in pure gauge/gravity in the preprint \cite{Tomboulis:1997gg}, concluding that the counter-terms are indeed local by a quite technical and formal proof, and without a concrete computation to support it. It would be worthwhile to cross-check this result and investigate the inclusion of the matter sector.

\item Still related to quantum corrections, Shapiro studied the dressed propagator\footnote{Computing the dressed propagator is a semiperturbative procedure, i.e. one computes the 1 particle irreducible (1PI) diagrams at a given order in perturbation theory, and then resum the whole chain of 1PI diagrams (nonperturbative ingredient), which is equivalent to resum the leading logarithms via the renormalization group in local QFT.} in superrenormalizable gravity, and he realized that it acquires an infinite tower of Lee-Wick poles, unless the remaining UV-divergences at 1-loop are canceled by some killer operators \cite{Shapiro:2015uxa}, like in the pure gravity $\partial^\infty$QFT's built in Ref.~\cite{Modesto:2014lga}. However, Modesto pointed out that these ``Shapiro ghosts'' are located outside the radius of convergence of the geometric series given by the infinite chain of 1PI diagrams \cite{Modesto:2021priv1}, i.e. they might just be spurious artifacts when trusting the expression of the dressed propagator outside its domain of definition. Therefore, the issue of ghost-freedom in $\partial^\infty$QFT at the level of the 1PI effective action fully deserves clarification elsewhere.

\item An important application could be the stabilization of scalar potentials under radiative corrections, with potentially important consequences for the gauge hierarchy problem \cite{Wilson:1970ag, Weinberg:1975gm, Gildener:1976ai, Susskind:1978ms, tHooft:1979rat, Veltman:1980mj, Kolda:2000wi} and vacuum stability \cite{Isidori:2001bm, Isidori:2007vm, Elias-Miro:2011sqh, Degrassi:2012ry, Buttazzo:2013uya, Rajantie:2016hkj, Salvio:2016mvj, Andreassen:2017rzq, Devoto:2022qen} in particle physics beyond the Standard Model \cite{Krasnikov:1987yj, Moffat:1990jj, Biswas:2014yia, Ghoshal:2017egr, Frasca:2020jbe, Nortier:2021six}.

\item Since there are essential singularities at complex infinity $\widetilde{\infty}$ in $\partial^\infty$QFT, the standard properties of scattering amplitudes need to be revisited \cite{Keltner:2015xda, Dona:2015tra, Talaganis:2016ovm, Buoninfante:2018gce, Tokuda:2019nqb, Koshelev:2021orf, Modesto:2021soh, Mo:2022szw, Buoninfante:2023dyd}.
\end{itemize}

\acknowledgments

Thanks to Raffaele T. D'Agnolo, Hermès Bélusca-Maïto, Pratik Chattopadhyay, Aldo Deandrea, Anish Ghoshal, Ratul Mahanta, Anupam Mazumdar, Nils Marion, Leonardo Modesto and Jean-Christophe Wallet for useful discussions and comments. Fig.~\ref{roots} was made with \texttt{Wolfram Mathematica}.

\appendix

\section{Conventions \& Notations}
\label{conventions}
\begin{itemize}[label=$\spadesuit$]
\item When not specified, the QFT conventions in this article are the same as in Peskin and Schroeder's textbook \cite{Peskin:1995ev}.

\item The $4$-dimensional Euclidean space is noted as $\mathbb{R}^{4}$ with the metric
\begin{equation}
\delta_{\mu\nu} = \text{diag}(+1, +1, +1, +1) \, .
\end{equation}
Euclidean quantities are written with the subscript $E$, e.g. the momentum $p_E$.

\item The $4$-dimensional Minkowski spacetime is noted as $\mathbb{R}^{1,3}$ with the metric
\begin{equation}
g_{\mu\nu} = \text{diag}(+1, -1, -1, -1) \, .
\end{equation}

\item The symbol $\cdot$ is used for either the pointwise product between fields or the scalar product of vectors.

\item Let $\mathbb{C}$ be the complex plane, the real and imaginary parts of $z \in \mathbb{C}$ are noted as $\mathfrak{R}(z)$ and $\mathfrak{I}(z)$, respectively.

\item Let $f(x)$ be a function on spacetime. Its Fourier transform in momentum space is written $\widetilde{f}(p)$.

\item One uses arrows on differential operators like $\overleftarrow{D}$ and $\overrightarrow{D}$ for operators acting on the function on the left and on the right, respectively.

\item Let $f(z)$ be an entire function on $\mathbb{C}$, the pseudodifferential operator $f(D)$ (where $D$ is some differential operator) is defined as
\begin{equation}
f(D) = \sum_{k=0}^{\infty} \dfrac{f^{(k)}(0)}{k!} \, D^k \, .
\end{equation}

\end{itemize}

\bibliographystyle{JHEP}

\providecommand{\href}[2]{#2}\begingroup\raggedright\endgroup

\end{document}